\documentclass[useAMS,usenatbib]{mn2e}
\usepackage{graphicx}
\usepackage{txfonts}
\usepackage{amssymb}
\usepackage{amsfonts}
\usepackage{times}
\usepackage{hyperref}
\usepackage{subfig}
\usepackage{mathbbol}
\pdfoutput=1

 \title[A new signature of primordial non-Gaussianities]
  {A new signature of primordial non-Gaussianities from the abundance of galaxy clusters}

\author[A. M. M. Trindade et al.]{A. M. M. Trindade$^{1,2}$\thanks{E-mail: Arlindo.Trindade@astro.up.pt} , P. P. Avelino$^{1,2}$ and P. T. P. Viana$^{1,2}$
\\
$^1$ Centro de Asrof\'{\i}sica da Universidade do Porto, Rua das Estrelas 687, 4150-762 Porto, Portugal\\
$^2$ Departamento de F\'{\i}sica e Astronomia da Faculdade de Ci\^encias da Universidade do Porto, Rua do Campo Alegre 687, 4169-007 Porto, Portugal}

\begin{document}

\date{Accepted 2012 May 16. Received 2012 April 22; in original form 2012 April 22}

\pagerange{\pageref{firstpage}--\pageref{lastpage}} \pubyear{2012}

\maketitle

\begin{abstract}
The evolution with time of the abundance of galaxy clusters is very sensitive to the statistical properties of the primordial density perturbations. It can thus be used  to probe small deviations from Gaussianity in the initial conditions. The characterization of such deviations would help distinguish between different inflationary scenarios, and provide us with information on physical processes which took place in the early Universe. We have found that when the information contained in the galaxy cluster counts is used to reconstruct the dark energy equation of state as a function of redshift, assuming erroneously that no primordial non-Gaussianities exist, an apparent evolution with time in the effective dark energy equation of state arises, characterized by the appearance of a clear discontinuity.
\end{abstract}
\begin{keywords}
Cosmology: large-scale structure of Universe -- Cosmology: dark energy
\end{keywords}

\section{Introduction}

One of the most fundamental predictions of the simplest standard, single field, slow-roll inflationary cosmology, is that the primordial 
density fluctuations, that seeded the formation of the large-scale structure we see today, were nearly Gaussian distributed 
(see e.g. \citealt{2003JCAP...10..003C,2003JHEP...05..013M,2005PhRvL..95l1302L,2005JCAP...06..003S,2007PhRvD..76h3004S}. 
Such prediction seems to be in good agreement with current observations of the cosmic microwave background anisotropies 
(e.g. \citealt{2008JCAP...08..031S}) and large-scale structure (e.g. \citealt{2011ApJS..192...18K}). Nevertheless, a significant, 
potentially observable level of non-Gaussianity may be produced in some inflationary models where any of the conditions that give 
rise to the standard single-field, slow-roll inflation fail.

The detection of primordial non-Gaussianities would decrease considerably the number of viable inflationary models, and it 
would give us an insight on key physical processes that took place in the early Universe. Such detection could be achieved 
through the statistical characterization of the properties of the large-scale structure, namely the bispectrum and/or 
trispectrum of the galaxy distribution (e.g. \citealt{2007PhRvD..76h3004S,2008ApJ...677L..77M}), or the determination 
of the evolution with time of the abundance of massive collapsed objects such as galaxy clusters 
(see e.g. \citealt{2000ApJ...541...10M,2000MNRAS.311..781R}). These form at high peaks of the density 
field $\delta \left(\textbf{x}\right)=\delta \rho/\rho$ and their number density as a function of redshift 
depends on the growth of structure, thus being sensitive to the dynamics and energy content of the Universe 
and to the statistical properties of the primordial density fluctuations.

In this work, we address the issue of how primordial non-Gaussianities may affect the determination of the 
effective dark energy equation of state $w$ using the evolution with time of the galaxy cluster abundance. 
Throughout, and unless stated otherwise, we consider our fiducial cosmological model to be a 
flat $\Lambda$CDM model with WMAP 7-year cosmological parameters (WMAP+BAO+$H_{0}$) \citep{2011ApJS..192...18K},  
namely, a Hubble constant, $H_{0}$, equal to $100h\,{\rm km}\,{\rm s}^{-1}\,{\rm Mpc}^{-1}$ with $h=0.704$, 
fractional densities of matter and baryons today of $\Omega_{m}=0.272$, $\Omega_{b}h^{2}=0.023$ respectively, 
a scalar spectral index, $n_{s}$, equal to 0.963, and we normalize the power spectrum so that $\sigma_{8}=0.809$.

\vspace{-0.4cm}
\section{Primordial Non-Gaussianity}\label{Section_2}

Primordial non-Gaussianity is commonly parametrized by the non-linear parameter $f_{NL}$ and may be written as follows \citep{2008JCAP...04..014L}, 

\begin{equation}
 \label{bispectrum_def_3}
B_{\zeta}\left(k_{1},k_{2},k_{3}\right)=\left(2\pi \right)^{4}f_{NL}\frac{\mathcal{P}_{\zeta}^{2}\left(K\right)}{\left(k_{1}k_{2}k_{3}\right)^{3}}
\mathcal{A}\left(k_{1},k_{2},k_{3}\right),
\end{equation}
where $K=k_{1}+k_{2}+k_{3}$, $\zeta$ is the primordial curvature perturbation and $\mathcal{A}$ is an auxiliar function that contains
 the shape of the bispectrum, $B_{\zeta}$, of $\zeta$. Further, $\mathcal{P}_{\zeta}\varpropto k^{n_{s}-1}$  is the dimensionless 
power-spectrum, while $k_i=|{\bf k}_i|$ with $\mathbf{k}_{i}$ being the wave vectors.
 In Fourier space, the functions $\mathcal{P}_{\zeta}$ and $B_{\zeta}$ are defined by means of the two and three-point correlation functions,

\begin{equation}
 \label{powerspectrum_def}
\left\langle \zeta_{\textbf{k}_{1}} \zeta_{\textbf{k}_{2}} \right\rangle=
\left(2\pi\right)^{3}\delta_{D}\left(\textbf{k}_{12}\right)
\frac{\left(2\pi\right)^{3}\mathcal{P}_{\zeta}\left(k_{1}\right)}{4\pi k_{1}^{3}}\,,
\end{equation}

\begin{equation}
 \label{bispecteum_def}
\left\langle \zeta_{\textbf{k}_{1}} \zeta_{\textbf{k}_{2}} \zeta_{\textbf{k}_{3}} \right\rangle=
\left(2\pi\right)^{3}\delta_{D}\left(\textbf{k}_{123}\right)
B_{\zeta}\left(k_{1},k_{2},k_{3}\right),
\end{equation}
where $\textbf{k}_{ij...n}\equiv \textbf{k}_{i}+\textbf{k}_{j}+...+\textbf{k}_{n}$.
Note that $f_{NL}$ may or may not be scale-dependent. In this work only the later case is considered.

The bispectrum of $\zeta$ is the lowest order statistics sensitive to non-Gaussian features. Depending on 
the underlying physical mechanism responsible for generating non-Gaussianities, different triangular 
configurations (shapes) will arise. There are broadly four classes of triangular shapes or, equivalently, 
four different bispectrum parametrizations can be found in the literature: Local, Equilateral, 
Folded and Orthogonal. Here we shall only consider the first two. They are defined as follows:
\begin{itemize}

\item[-] \textbf{Local shape}: It arises from multi-field, inhomogeneous reheating, curvaton and 
ekpyrotic models. Mathematically, the Local shape can be characterized by a simple 
Taylor expansion around the Gaussian curvature perturbation, $\zeta_{G}$, \citep{1990PhRvD..42.3936S,2001PhRvD..63f3002K},
\begin{equation}
\label{curvature_taylor_expansion}
\zeta\left(\mathbf{x}\right)=\zeta_{G}\left(\mathbf{x}\right)+\frac{3}{5}f^{local}_{NL} \left(\zeta_{G}^{2}\left(\mathbf{x}\right)-\langle\zeta_{G}^{2}
\left(\mathbf{x}\right)\rangle\right)\,.
\end{equation}
The WMAP 7-year estimate for $f_{NL}^{local}$ is \citep{2011ApJS..192...18K} 
\begin{equation}
\label{local_estimate}
f_{NL}^{local}=32\pm 21 \left( 68\% \,C.L.\right)\,.
\end{equation}
The bispectrum for this shape can be derived using Eq. (\ref{curvature_taylor_expansion}) and it is given by \citep{2008JCAP...04..014L,2010CQGra..27l4010K}
\begin{eqnarray}
\label{local_form}
\nonumber \mathcal{A}_{local}=\frac{3}{10}K^{-2\left(n_{s}-1\right)}\left[k_{1}^{3}\left(k_{2}k_{3}\right)^{n_{s}-1}+ \right.\\ 
  \left. k_{2}^{3}\left(k_{1}k_{3}\right)^{n_{s}-1}+k_{3}^{3}\left(k_{1}k_{2}\right)^{n_{s}-1}\right]\,.
\end{eqnarray}
This quantity is maximized for the so-called squeezed triangle configuration, i.e. $k_{3}\ll k_{2}\approx k_{1}$. 

\item[-] \textbf{Equilateral shape}: It is characteristic of inflationary models where scalar fields have a non-canonical kinetic term (for example Dirac-Born-Infield inflation, see \citealt{2004PhRvD..70l3505A}). The mathematical expression for the Equilateral shape is \citep{2008JCAP...04..014L,2010CQGra..27l4010K}
\begin{eqnarray}
\label{equilateral_form}
\nonumber \mathcal{A}_{equi}=\frac{9}{10}K^{-2\left(n_{s}-1\right)}\left[-k_{1}^{3}\left(k_{2}k_{3}\right)^{n_{s}-1}+ \mathit{perm.} 
\right. \\  \left. -2\left(k_{1}k_{2}k_{3}\right)^{1+2\left(n_{s}-1\right)/3} + \right. 
\left. k_{1}^{2+\left(n_{s}-1\right)/3}k_{2}^{1+2\left(n_{s}-1\right)/3}k_{3}^{n_{s}-1} + \mathit{perm.}  \right]\,.
\end{eqnarray}
This quantity reaches a  maximum at $k_{1}\approx k_{2}\approx k_{3}$. Constraints from WMAP 7-year set the
 level of non-Gaussianity for this shape at \citep{2011ApJS..192...18K}
\begin{equation}
\label{equilateral_estimate}
f_{NL}^{equi}=26\pm 140 \left( 68\% \,C.L.\right)\,.
\end{equation}
\end{itemize}

As mentioned before, the abundance of rare objects such as galaxy clusters holds relevant information that can 
be used to probe the initial conditions. For such information to be of use, the statistics of the density 
perturbation, $\delta_{R}$, smoothed on a scale $R$, have to be characterized. However, we have previously 
defined the non-Gaussianity in the primordial curvature perturbation, $\zeta$, rather than in the smoothed 
linear density field. The relation between $\zeta$ and the linear perturbation to the matter density today smoothed on a scale $R$ is given by,
\begin{equation}
 \label{smoothed_density}
\delta_{R}\left(\textbf{k},z\right)=D\left(z\right)W(k,R) \mathcal{M}(k)T(k)\zeta \left(\textbf{k},z\right),
\end{equation}
where 
\begin{equation}
 \mathcal{M}(k)=\frac{2}{5}\frac{1}{\Omega_{m}}\frac{c^{2}}{H_{0}^{2}}k^{2},
\label{MM1}
\end{equation}
with $D\left(z\right)$ being the linear growth factor \citep{2009ApJS..180..330K}, $T\left(k\right)$ is the transfer function 
adopted from \cite{1986ApJ...304...15B}, and $W\left(k,R\right)$ is the smoothing top-hat window. We use the shape parameter 
given by \cite{1995ApJS..100..281S}, ${\Gamma=\Omega_{m}h\exp \left[-\Omega_{b}\left(1+\sqrt{2h}/\Omega_{m}\right)\right]}$.

Using Eq. (\ref{smoothed_density}) and the definitions given in Eqs. (\ref{powerspectrum_def}) and (\ref{bispecteum_def}), one may compute the variance 
\begin{eqnarray}
\label{2point_function}
\sigma^{2}\left(R\right)=\delta_{R}^{2} &=&\int \frac{d^{3}\mathbf{k_{1}}}{\left(2\pi\right)^3}\int \frac{d^{3}\mathbf{k_{2}}}{\left(2\pi\right)^3}
\mathcal{F}_{1} \mathcal{F}_{2} \langle \zeta_{1} \zeta_{2}\rangle  = \nonumber \\
&=&\int_{0}^{\infty} \frac{dk}{k}\mathcal{F}^{2}\left(k\right) \mathcal{P}_{\zeta}\left(k\right)\,,
\end{eqnarray}
and the three-point function for the smoothed density field \citep{2008JCAP...04..014L}
\begin{eqnarray}
 \label{3point_corr}
 \langle \delta_{R}^{3}\rangle=f_{NL}\int \frac{d^{3}\mathbf{k_{1}}}{\left(2\pi\right)^{3}}\int \frac{d^{3}\mathbf{k_{2}}}{\left(2\pi\right)^{3}} 
\int \frac{d^{3}\mathbf{k_{2}}}{\left(2\pi\right)^{3}}
\mathcal{F}_{1} \mathcal{F}_{2} \mathcal{F}_{3}\langle \zeta_{1} \zeta_{2} \zeta_{3}\rangle 
\nonumber \\
=\int \frac{d^{3}\mathbf{k_{1}}}{\left(2\pi\right)^{3}}\int \frac{d^{3}\mathbf{k_{2}}}{\left(2\pi\right)^{3}}
\mathcal{F}_{1} \mathcal{F}_{2} \mathcal{F}_{12}\left(2\pi \right)^{4}\left(\mathcal{P}_{\zeta}\left(K\right)\right)^{2}
\frac{\mathcal{A}\left(k_{1},k_{2},k_{12}\right)}{\left( k_{1}k_{2}k_{12}\right)^{3}}\,,
\end{eqnarray}
with $\zeta_{i}\equiv \zeta\left(\mathbf{k_{i}}\right)$, $\mathcal{F}_{i}\equiv W\left(k_{i},R\right)\mathcal{M}\left( k_{i}\right)T\left(k_{i}\right)$, $K=k_{1}+k_{2}+k_{3}$ and $k_{12}=\sqrt{k_{1}^{2}+k_{2}^{2}+2\mathbf{k}_{1}\cdot\mathbf{k}_{2}}$ 
(here a dot represents the scalar product). Eqs. (\ref{2point_function}) and (\ref{3point_corr}) will be of special relevance in the next section, since in order to incorporate non-Gaussian initial conditions
in the prediction of rare objects, one has to derive a non-Gaussian probability density function (PDF) for the smoothed density field, $\delta_{R}$. This can be done by using a mathematical procedure that enables us to construct the PDF from its comulants (see \citealt{2008JCAP...04..014L,2008ApJ...677L..77M} for details).

\section{The dark energy equation of state from the galaxy cluster abundance}

\subsection{Halo Mass Function}
\begin{figure*}
\centering
\subfloat[]{\includegraphics[scale=0.3]{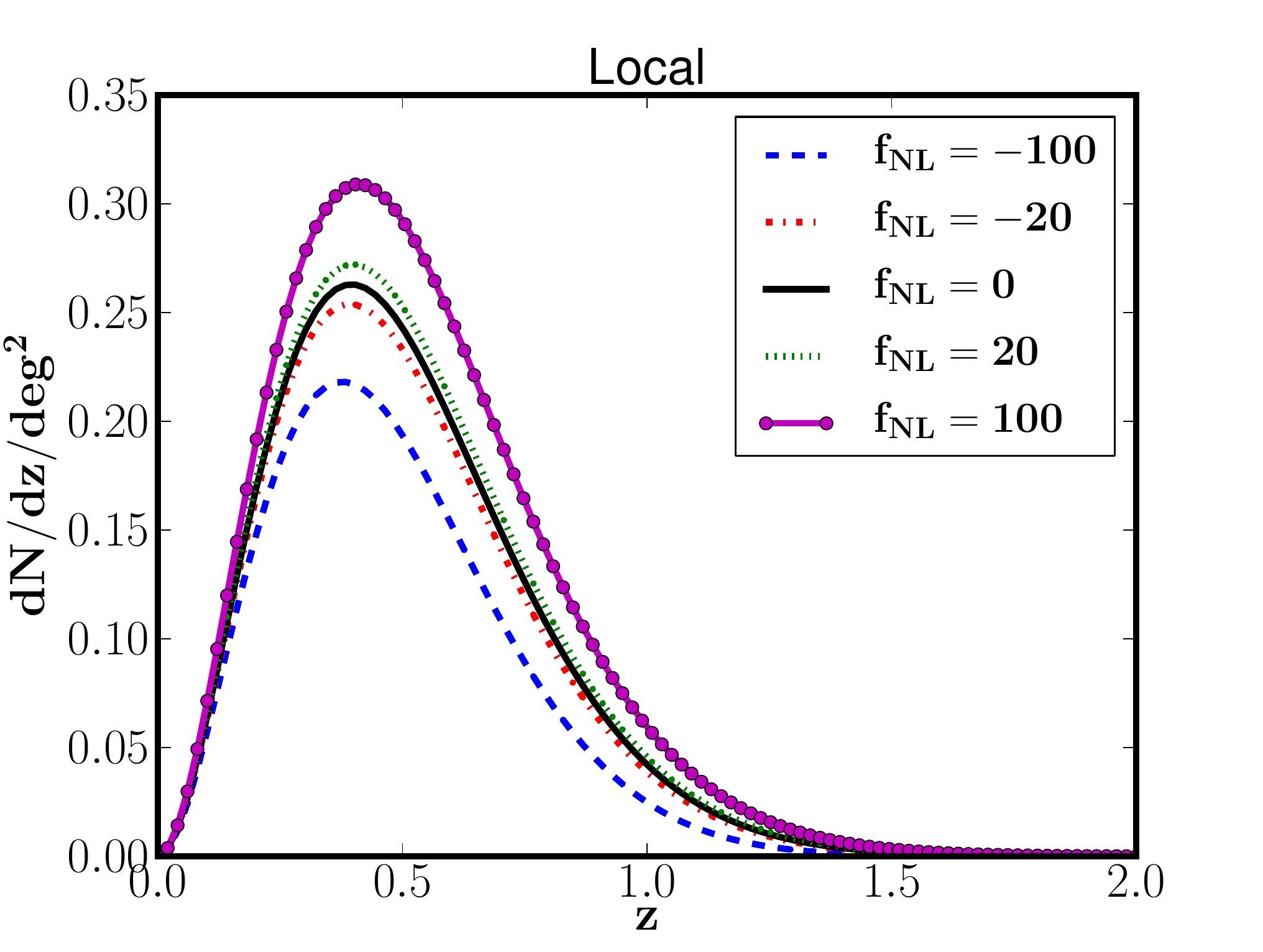}\label{Figure1a}}
\subfloat[]{\hspace{-0.4cm}\includegraphics[scale=0.3]{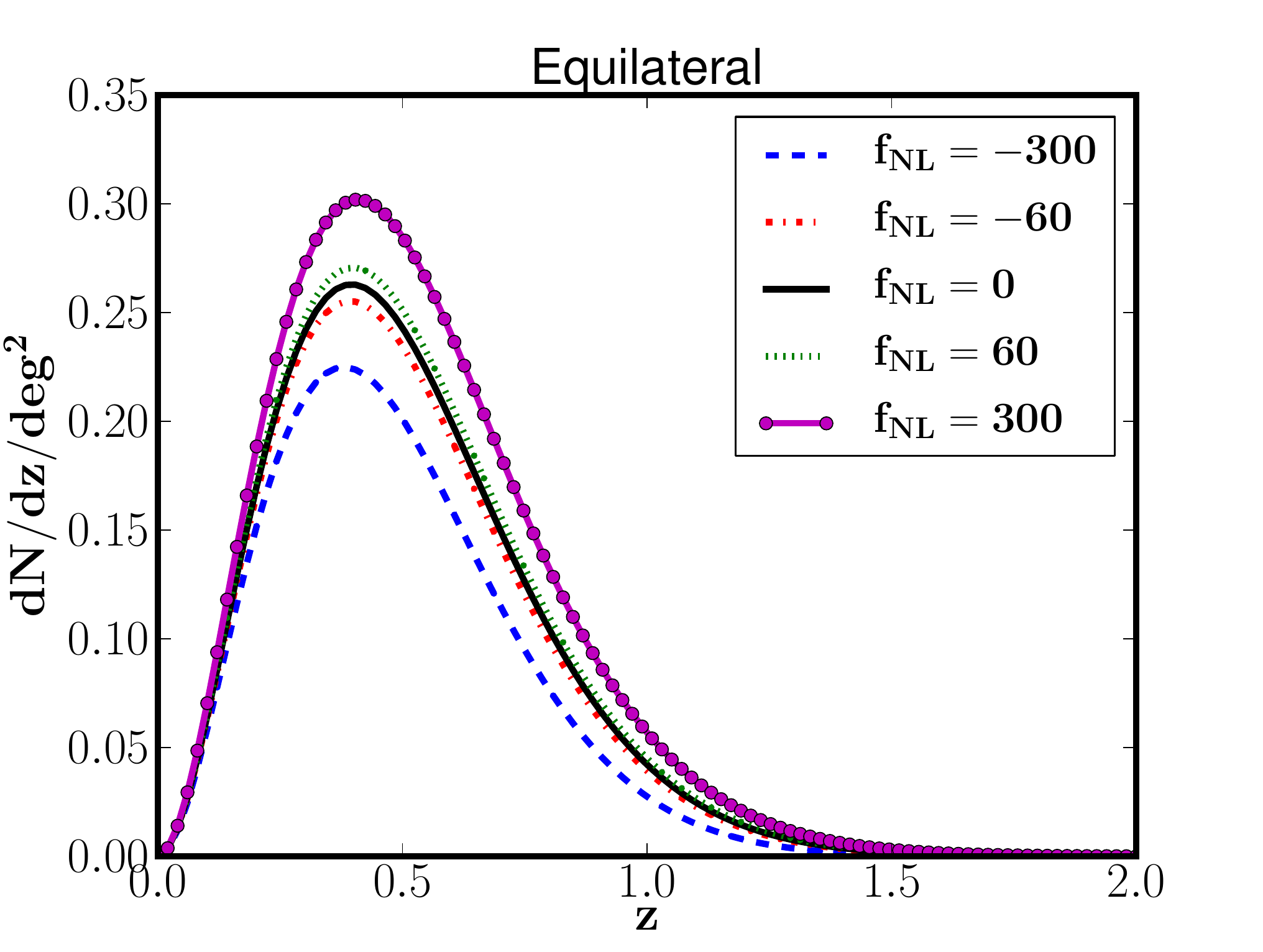}\label{Figure1b}}
\subfloat[]{\hspace{-0.4cm}\includegraphics[scale=0.3]{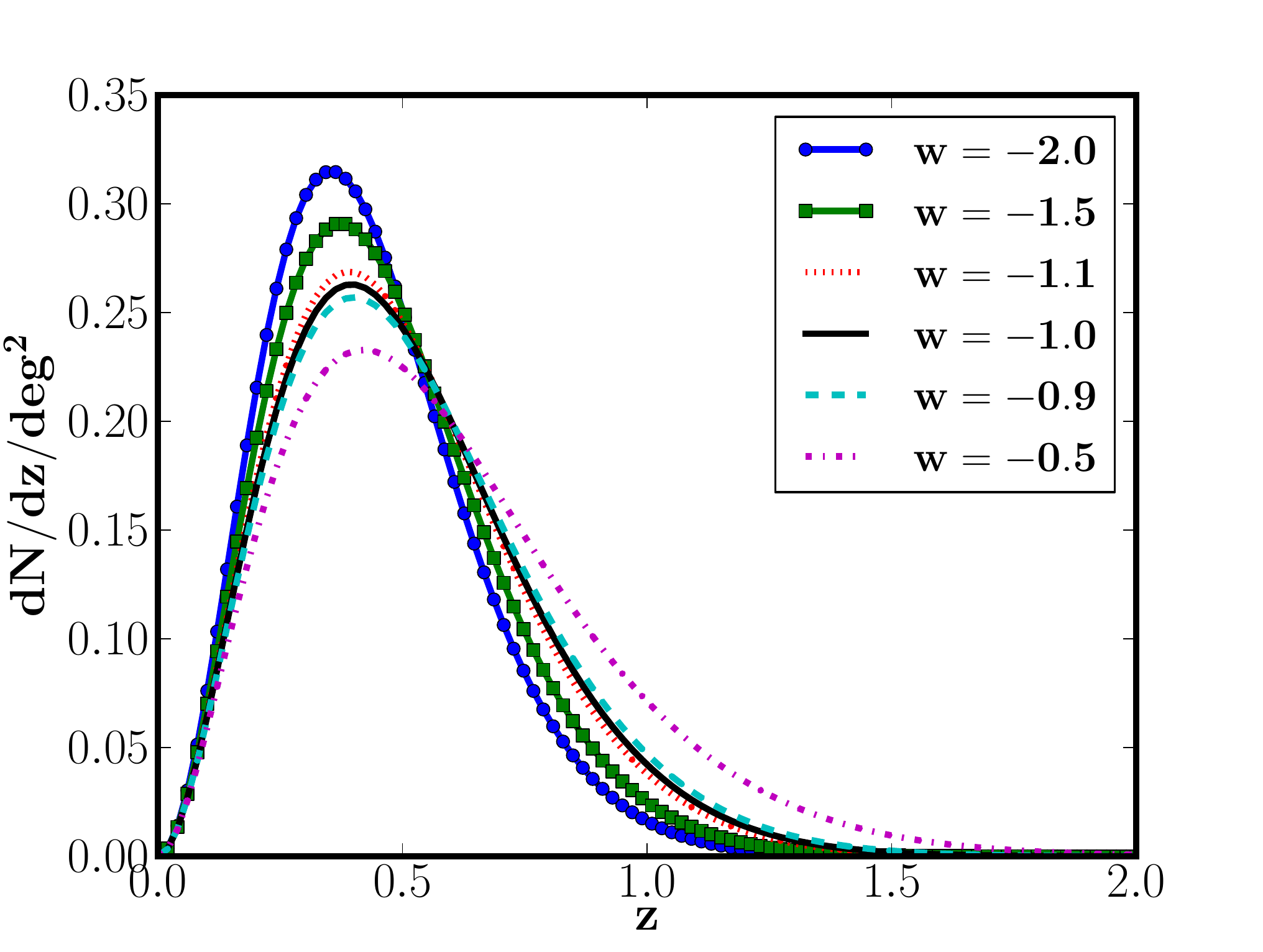}\label{Figure1c}}

\caption{ The number of galaxy clusters per unit of redshift per square degree with mass
 $M>M_{lim}=5\times10^{14}\, h^{-1}M_{\odot}$, considering $w=-1$ and different levels of non-Gaussianity 
 for (a) Local and (b) Equilateral parametrizations.  Panel $\left(c\right)$ shows 
 the effect that a change on a constant dark energy equation of state parameter, $w$, has on the number of clusters per unit of redshift per square degree with Gaussian initial conditions ($f_{NL}=0$).}
\label{Figure1}
\end{figure*}

The comoving number density of virialized halos per unit of volume at a given redshift $z$, $dn/dM\left(z,M\right)$, with a mass, $M$, in the range $[M, M + dM]$ is called the mass function. In the presence of non-Gaussian initial conditions, the mass function has been estimated using extensions of the Press-Schechter (PS) formalism \citep{1974ApJ...187..425P}. This formalism asserts that the fraction of matter ending up in objects of mass $M$ is proportional to the probability that the density fluctuations smoothed on the scale $R=\left(3M/4 \pi \overline{\rho}\right)^{\frac{1}{3}}$, and above a certain threshold value, $\delta_{c}$, can be written as,
 \begin{eqnarray}
  \label{mass_function_def}
 \frac{dn}{dM}\left(z,M\right)=-2\frac{\overline{\rho}}{M}\frac{d}{dM}\left[\int_{\delta_{c}\left(z\right)/ \sigma_{M}}^{\infty} d\nu P\left(\nu,M\right)\right]
 \end{eqnarray}
where $\overline{\rho}$, $\sigma_{M}$, $\delta_{c}$ and $P\left(\nu,M\right)$ are respectively, the comoving mass density, the root square of the variance of the density perturbation in spheres of radius $R$, the critical overdensity in the spherical collapse model and the PDF of the smoothed density field. The redshift dependence of the threshold for spherical collapse as been incorporated in $\delta_{c}\left(z\right)=1.686D\left(0\right) D^{-1}\left(z\right)$. For Gaussian initial conditions the mass function acquires the following form,
\begin{equation}
\label{PS_MF}
\frac{dn}{dM}\left(M,z\right)=-\sqrt{\frac{2}{\pi}}\frac{\bar{\rho}}{M^{2}}\frac{\delta_{c}\left(z\right)}{\sigma_{M}}
\frac{d\ln\,\sigma_{M}}{d\ln\, M} \mathit{e}^{-\delta_{c}^{2}\left(z\right)/\left(2\sigma_{M}^{2}\right)}\,.
\end{equation}
The cosmological parameters enter Eq. (\ref{PS_MF}) essentially through the variance and the linear growth factor, as well as,  through the critical density contrast $\delta_{c}\left(z\right)$.

There are several prescriptions for changing Eq. (\ref{PS_MF}) in the presence of non-Gaussian initial conditions \citep{1998ApJ...494..479C,2000ApJ...532....1R,2000MNRAS.314..354A,2000ASPC..200..408F,2000ApJ...541...10M}. Here we will adopt that which has been proposed by \cite{2008JCAP...04..014L}, and which has been shown to provide a good fit to results from N-body simulations (see \citealt{2010JCAP...10..022W} and references therein),
\begin{eqnarray}
\label{ng_mass_function}
\frac{dn}{dM}\left(M,z\right)=-\sqrt{\frac{2}{\pi}}\frac{\overline{\rho}}{M^{2}}\mathit{e}^{-\delta_{c}^{2}\left(z\right)/2\sigma_{M}^{2}}\left[\frac{d\ln \sigma_{M}}{d\ln M}
\left(\frac{\delta_{c}\left(z\right)}{\sigma_{M}} + \frac{S_{3M}\sigma_{M}}{6} \right. \right. \nonumber \\ 
\left. \left. \times \left(\frac{\delta_{c}^{4}\left(z\right)}{\sigma_{M}^{4}} -2\frac{\delta_{c}^{2}\left(z\right)}{\sigma_{M}^{2}}-1 
 \right) \right) +\frac{1}{6}\frac{dS_{3M}}{d\ln M}\sigma_{M}\left(\frac{\delta_{c}^{2}\left(z\right)}{\sigma_{M}^{2}}-1\right)\right]\,,
\end{eqnarray}
where $S_{3M}=\langle\delta_{M}^{3}\rangle/\langle\delta_{M}^{2}\rangle^{2}\propto f_{NL}$ is the skewness of the smoothed density field. If $f_{NL}=0$, then $S_{3M}=0$ and Eq. (\ref{ng_mass_function}) reduces to the Gaussian mass function. 

Numerical simulations have shown that the PS form of the mass function under-predicts the abundance of high-mass objects and over-predicts low-mass ones. Therefore, to be in better agreement with the results of numerical simulations, we follow \cite{2008JCAP...04..014L} and model the departures from Gaussianity using the mass function suggested by \cite{2001MNRAS.323....1S}. The non-Gaussian mass function then becomes,
\begin{equation}
\label{mass_function_new}
\frac{dn_{NG}}{dM}\left(z,M,f_{NL}\right)=\frac{dn_{ST}}{dM}\frac{dn_{PS}/dM(z,M,f_{NL})}{dn_{PS}/dM(z,M,f_{NL}=0)}\,.
\end{equation}

The Seth-Tormen mass function has been calibrated using numerical simulations with Gaussian initial conditions and it is given by \citep{2001MNRAS.323....1S},
\begin{eqnarray}
\label{ST_mass_function}
\frac{dn_{ST}}{dM}\left(z,M\right)&=&-\sqrt{\frac{2a}{\pi}}A\left(1+\left(\frac{a\delta_{c}}{\sigma^{2}}\right)^{-p}\right)
\frac{\bar{\rho}}{M^{2}}\frac{\delta_{c}\left(z\right)}{\sigma_{M}} \nonumber \\
 &\times& \frac{d\ln\,\sigma_{M}}{d\ln\, M} \mathit{e}^{-a\delta_{c}^{2}\left(z\right)/\left(2\sigma_{M}^{2}\right)}
\end{eqnarray}
with $a=0.707$, $A=0.322184$ and $p=0.3$.

Given Eqs. (\ref{mass_function_new}) and (\ref{ST_mass_function}), one may now compute the number of clusters of galaxies per unit of redshift above a certain mass threshold, $M_{lim}$,
\begin{equation}
\label{Cluster_abundance}
\frac{dN}{dz}\left(z,M>M_{lim}\right)=f_{sky}\frac{dV}{dz}\left(z\right)\int_{M_{lim}}^{\infty}dM\frac{dn}{dM}\left(z,M,f_{NL}\right),
\end{equation}
where $f_{sky}$ is the fraction of sky being observed in the survey and $dV/dz(z)$ is the comoving volume element which, for a flat cosmology, is given by
\begin{equation}
  \label{volume}
\frac{dV}{dz}\left(z\right)=4\pi\, \chi\left(z\right) \frac{d \chi}{dz}\left(z\right),
 \end{equation}
with $\chi \left(z\right)$ being the comoving radial distance.

Figures \ref{Figure1a} and \ref{Figure1b} illustrate the impact of the primordial non-Gaussianities 
on the number density of galaxy clusters with mass threshold $M>M_{lim}=5\times10^{14}\, h^{-1}M_{\odot}$ 
and $w=-1$, as a function of redshift, for the Local and Equilateral parametrizations, respectively. 
Figure \ref{Figure1c} shows how the cluster number density is affected when a constant dark energy 
equation of state parameter $w$ is changed for the same mass threshold, always assuming 
Gaussian initial conditions. The effect of $f_{NL}$ and $w$ on the abundance of galaxy clusters is 
quite different. On one hand, increasing $w$ above $w=-1$ flattens the slope of the cluster abundance 
above $z\approx 0.5$, which translates in an increase in the number of high-z clusters. On the other 
hand, changes in $f_{NL}$ modify the cluster abundance more uniformly in redshift. The difference in 
behaviour occurs because $w$ affects both the volume factor and the mass function, while $f_{NL}$ 
changes only the tail of the distribution of the density perturbations, thus modifying just the mass function. 

\begin{figure*}
\centering
\subfloat[]{\includegraphics[scale=0.39]{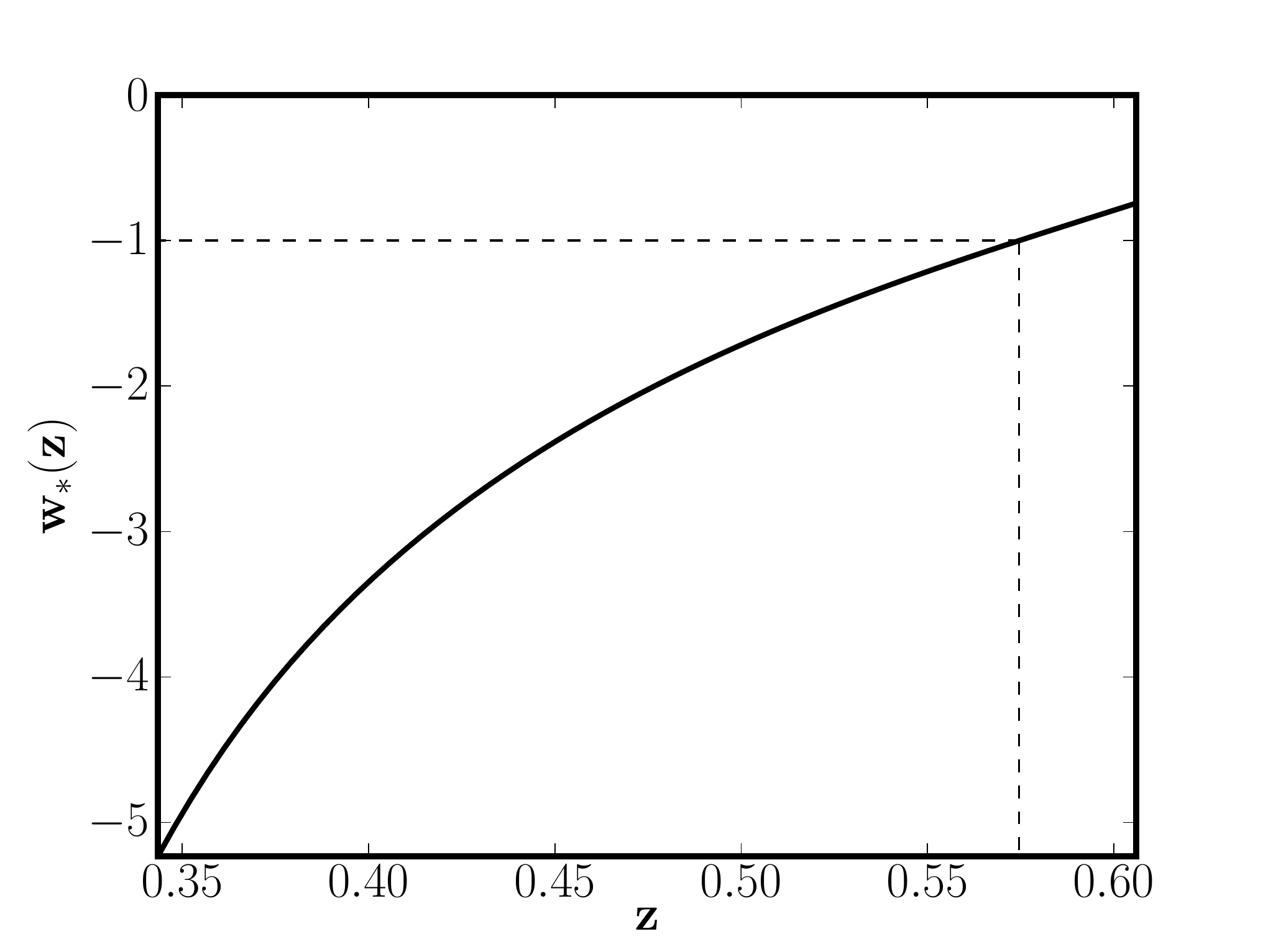}\label{Figure2a}}
\subfloat[]{\includegraphics[scale=0.39]{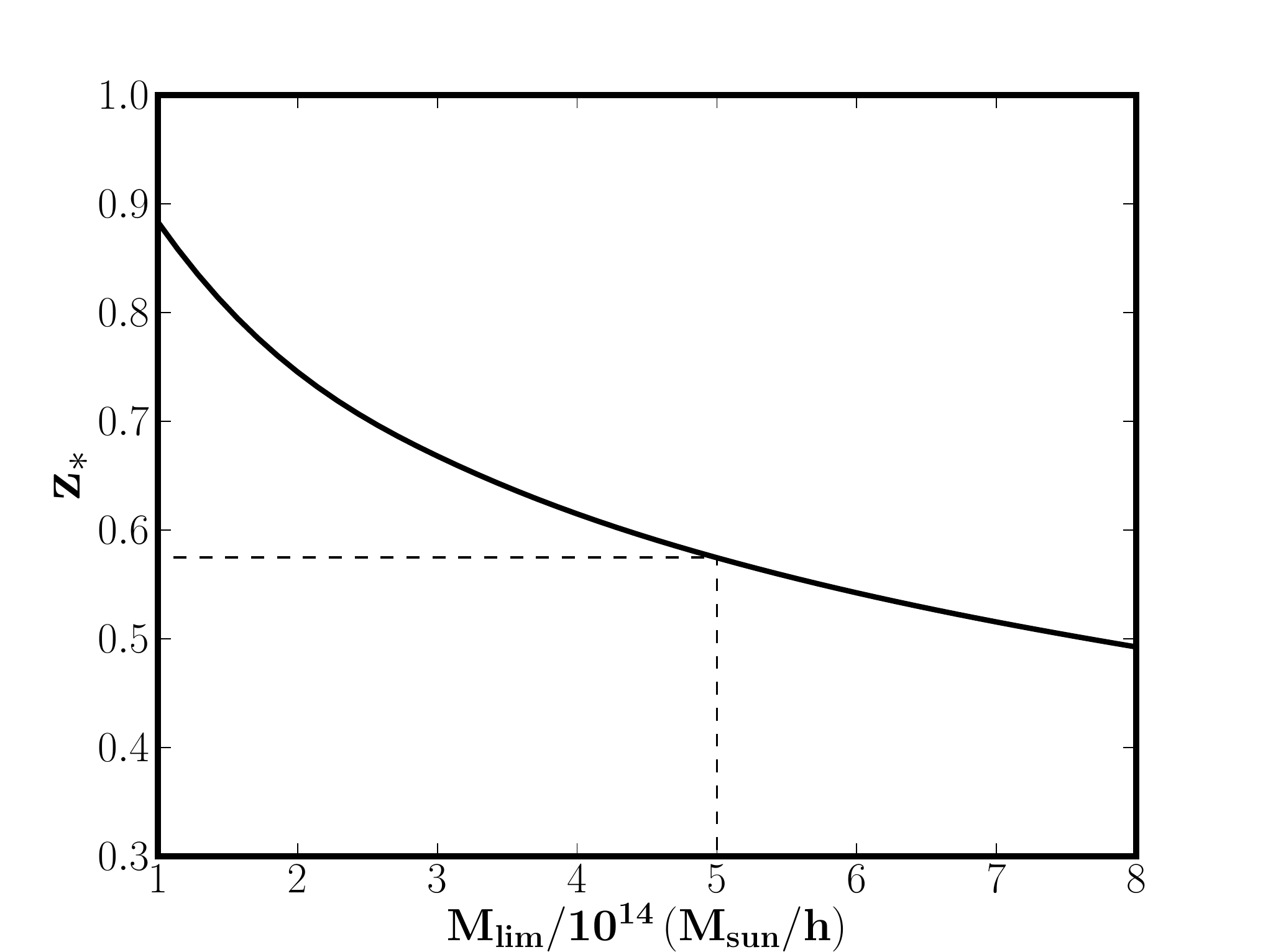}\label{Figure2b}}\\
\subfloat[]{\includegraphics[scale=0.39]{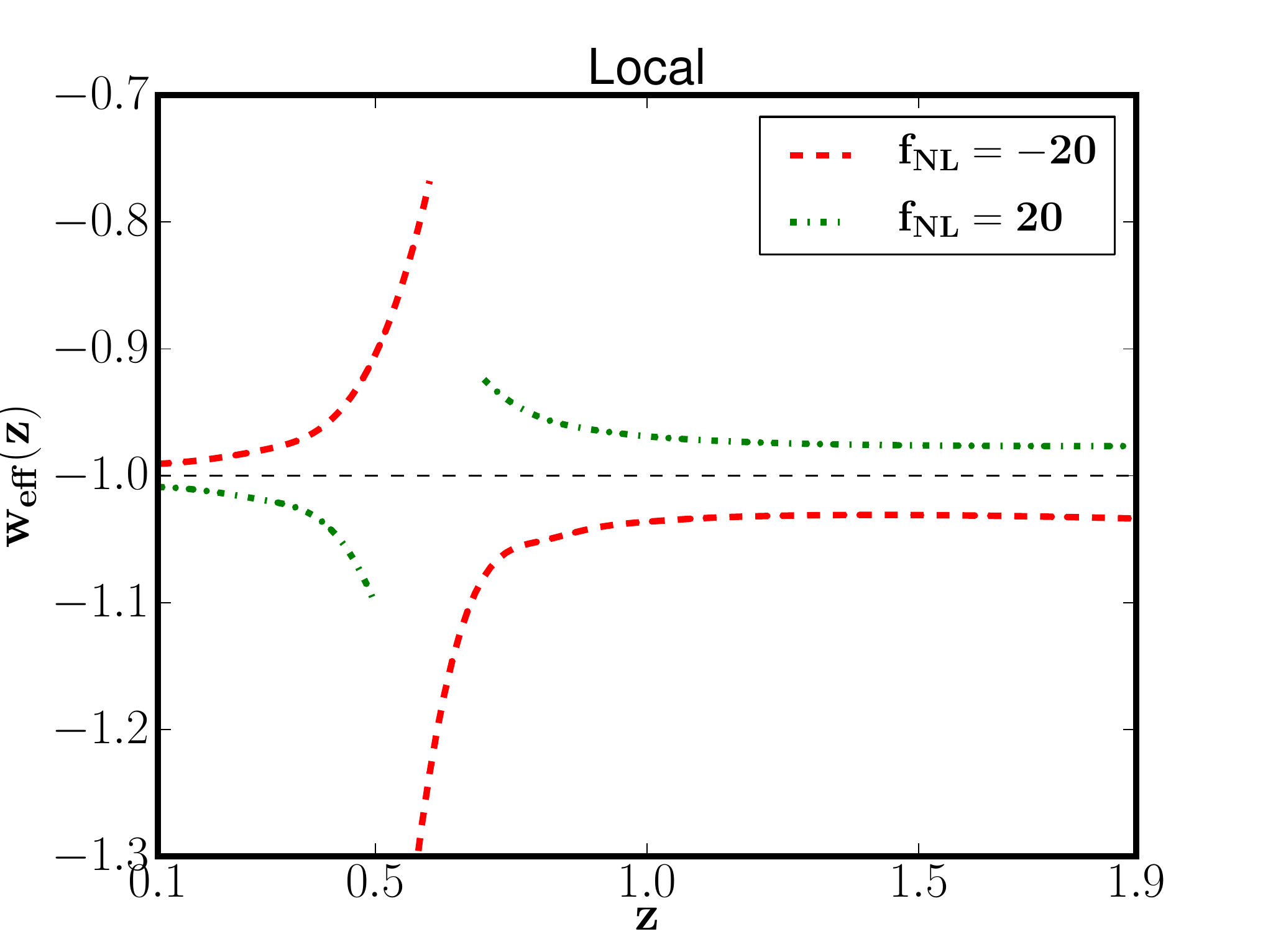}\label{Figure2c}}
\subfloat[]{\includegraphics[scale=0.39]{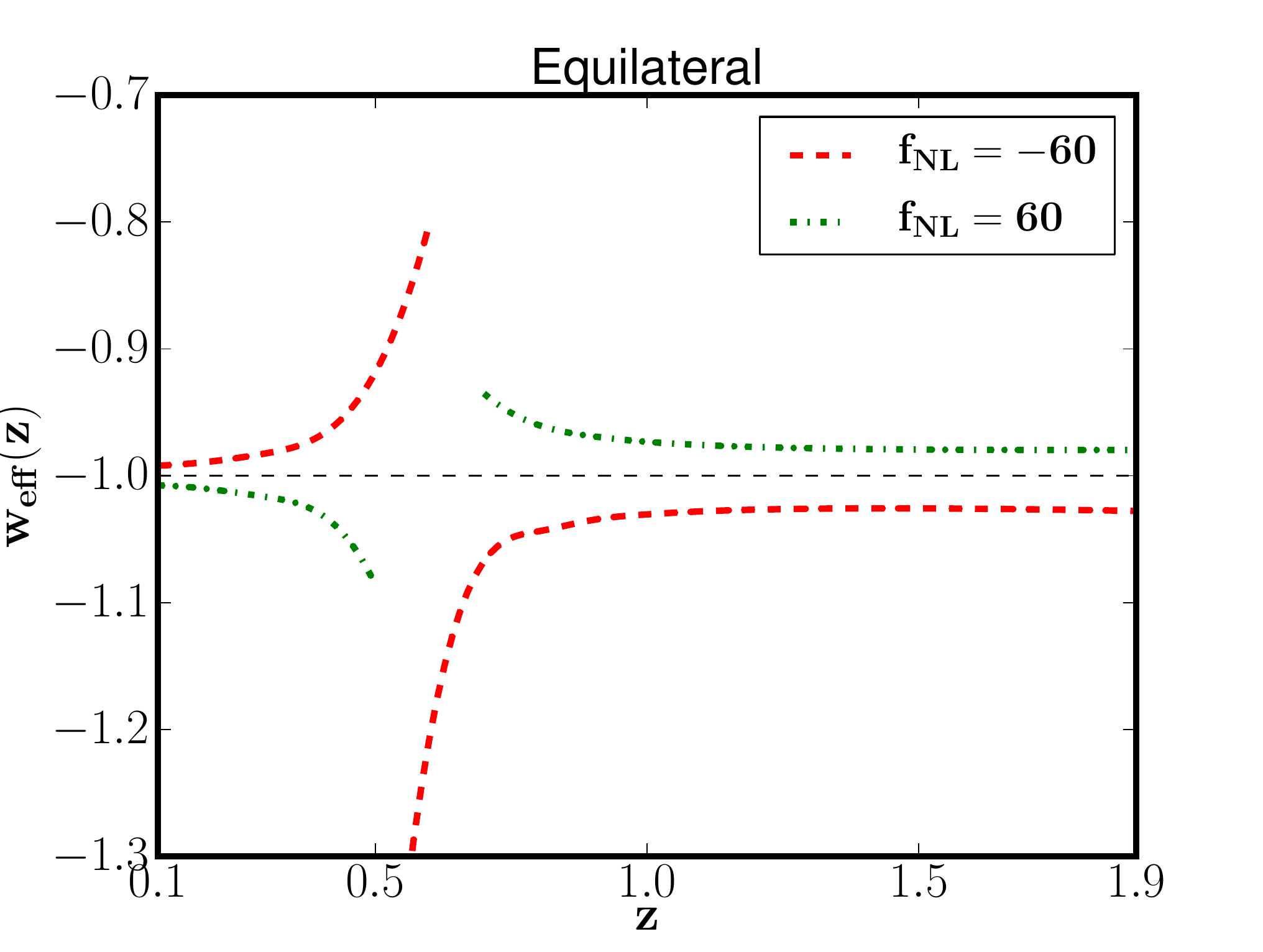}\label{Figure2d}}

\caption{Figure \ref{Figure2a} shows the value of $w_{eff}$ that maximizes the abundance of clusters as a function of redshift, with Gaussian initial conditions and $M_{lim}=5\times10^{14}\, h^{-1}M_{\odot}$ (the vertical dotted line corresponds to the redshift $z_{*} \sim 0.575$ where $w_{*}=-1$). Figure \ref{Figure2b} shows the dependence of the of the value of $z_*$ on the mass threshold $M_{lim}$. Figures \ref{Figure2c} and \ref{Figure2d} show the reconstructed effective dark energy equation of state $w_{eff}$ for $M>M_{lim}=5\times10^{14}\, h^{-1}M_{\odot}$ and different values of $f_{NL}$ (Local  and Equilateral parametrizations, respectively).}
\label{Figure3} 
\end{figure*}

\begin{table*}
\centering
\begin{tabular}{cccccc}
\hline
\hline Model &  Local &  Equilateral \\
\hline
\hline $f_{NL}$ & -20 \hspace{0.5cm} 20 & -60 \hspace{0.5cm} 60 \\ 
\hline $\sigma_{8,NG}$ &  0.812 \hspace{0.2cm} 0.806& 0.813 \hspace{0.2cm} 0.807  \\ 
\hline 
\hline
\end{tabular}
\caption{ The computed  $\sigma_{8}$ for the Local parametrization and Equilateral parametrization, obtained by demanding the present-day number density of galaxy clusters is recovered. }
\label{Table1}
\end{table*}

\vspace{-0.5cm}
\subsection{Estimation of w$_{eff}$}
\label{w_eff_reconstruction}

Figures \ref{Figure1a}, \ref{Figure1b} and \ref{Figure1c}, suggest that the redshift evolution of the number density of galaxy clusters in 
non-Gaussian models could be wrongly taken to be the result of an effective dark energy equation of state different from the real one, 
under the assumption of Gaussian initial conditions. In order to test this hypothesis, we have generated mock catalogues with the 
expected redshift evolution of the cluster number density for different non-Gaussian initial conditions in bins of redshift with width 
$\Delta z=0.1$ up to redshift $2$,assuming a mass threshold of $M_{lim}=5\times10^{14}h^{-1}M_{\odot}$ and a nearly full sky 
survey area of 40,000 square degrees. We further consider $f_{NL}^{Local}=\left(-20,20\right)$ and $f_{NL}^{Equil.}=\left(-60,60\right)$, 
respectively for theLocal and Equilateral parametrizations mentioned in section \ref{Section_2}. The reason for choosing these specific 
values for $f_{NL}$ will become clear in the next subsection.

Having generated the mock catalogues with the expected redshift distribution of the number density of galaxy clusters with non-Gaussian 
initial conditions, we then computed an effective dark energy equation of state, $w_{eff}$, using Eq. (\ref{Cluster_abundance}) with 
$f_{NL}=0$, that mimics the distribution of the number of galaxy clusters in the presence of non-Gaussian initial conditions at the 
\textit{i-th} redshift bin, by solving the following equation,

\begin{eqnarray}
 \label{w_determination}
N^{G}_{i}&=&\int^{z_{i}+\Delta z/2}_{z_{i}-\Delta z/2} \frac{dN}{dz}\left(z,w=w_{eff},f_{NL}=0\right)dz=\nonumber \\
&=&\int^{z_{i}+\Delta z/2}_{z_{i}-\Delta z/2} \frac{dN}{dz}\left(z,w=-1,f_{NL}\neq0\right)dz=N^{NG}_{i}
\end{eqnarray}
with respect to $w_{eff}$ for each redshift bean. Thus, the effective dark energy equation of state, $w_{eff}(z)$, is defined as the value 
of $w$ which reproduces the number of clusters of the mock non-Gaussian catalogue in \textit{i-th} redshift bin.

The normalization of the power-spectrum, for models with non-Gaussian initial conditions, 
was done by demanding the present-day number density of galaxy clusters in the concordance model 
($\Lambda$CDM cosmology with Gaussian initial conditions and $\sigma_{8}=0.809$) is recovered. 
Table \ref{Table1} shows the computed $\sigma_{8}$ for different values of $f_{NL}$ and Local and Equilateral parametrizations.

Figures \ref{Figure2c} and \ref{Figure2d} show the reconstructed dark energy equation of state, $w_{eff}$, 
as a function of the redshift for $M_{lim}=5\times10^{14}\, h^{-1}M_{\odot}$ and different values of $f_{NL}$ (Local and 
Equilateral parametrizations, respectively). At low redshift, our reconstructed $w_{eff}$ is very close to our 
fiducial $w=-1$. But, as we move towards higher redshifts, our computed $w_{eff}$ deviates from $-1$, 
with this effect being more evident for higher values of $f_{NL}$ in both  parametrizations. Further, 
for $f_{NL}>0$ there is a redshift interval where $w_{eff}$, is undefined, which widens with increasing $f_{NL}$. 
For $M_{lim}=5\times10^{14}\, h^{-1}M_{\odot}$ this happens in a redshift range centered at the redshift 
$z_{*} \sim 0.575$, at which the value of $w_{eff}$, which we will call $w_{*}$, that maximizes the cluster 
abundance is equal to $-1$  (see Figure \ref{Figure2a}). In this interval, there is no $w_{eff}$ that solves 
Eq. (\ref{w_determination}), since the product of the comoving volume with the integral of the mass function, 
when assuming Gaussian initial conditions, is always smaller 
than the same quantity for non-Gaussian initial conditions, i.e. $f_{NL}\neq0$. If our fiducial cosmological model 
had a different value for $w$, then the discontinuity would appear at the redshift at which $w=w_{*}$. On the other 
hand, for models with $f_{NL}<0$, we have also a discontinous $w_{eff}$ but there are two values of $w_{eff}$ capable of 
reproducing the number of clusters in non-Gaussian models in a small redshift interval centered at $z_{\star}$. 
The dependence of the value of $z_*$ on the mass threshold $M_{lim}$ is shown in figure \ref{Figure2b}.
\vspace{-0.5cm}
\subsection{Estimation of w$_{eff}$ with statistical uncertainties}

The observational estimation of the number density of galaxy clusters is affected by two sources of statistical 
uncertainty: the shot-noise and the cosmic variance (e.g. \citealt{2011A&A...536..A95V}). The statistical uncertainty 
associated with the former increases, for example, with the cluster mass threshold, as clusters then become 
more rare, while the statistical uncertainty associated with the later increases, for example, as the cosmic 
volume surveyed gets smaller. However, assuming primordial Gaussian density perturbations, and for a cluster 
mass threshold above $5\times10^{14}h^{-1}M_{\odot}$, it has been shown that the magnitude of the contribution 
of the cosmic variance to the statistical uncertainty, in the observed number density of galaxy clusters, 
is always at least an order of magnitude smaller than the contribution due to shot-noise, almost independently 
of the surveyed sky area (e.g. see figures 5 and 9 of \citealt{2011A&A...536..A95V}). 

The existence of statistical uncertainties in the observed number density of galaxy clusters will to some extent mask the 
apparent discontinuity on the evolution of $w_{eff}$ described in the previous section, which could be used to identify 
the presence of non-Gaussian density perturbations. In order to determine the impact of such uncertainties, we will 
re-estimate $w_{eff}$ in similar fashion to what was done in the previous section, but with the inclusion of shot-noise, 
$\sigma_{N}=\sqrt{N_{i}}$, at the $1\sigma$ level in each redshift bin. We are able to neglect the contribution from cosmic 
variance by setting the cluster mass threshold to $5\times10^{14}h^{-1}M_{\odot}$, and noting that the assumed level 
of non-Gaussianity is relatively small, not affecting much the average number density of galaxy clusters with respect to the 
Gaussian case (see Fig. 1).

\begin{figure*}
\centering
\includegraphics[scale=0.42]{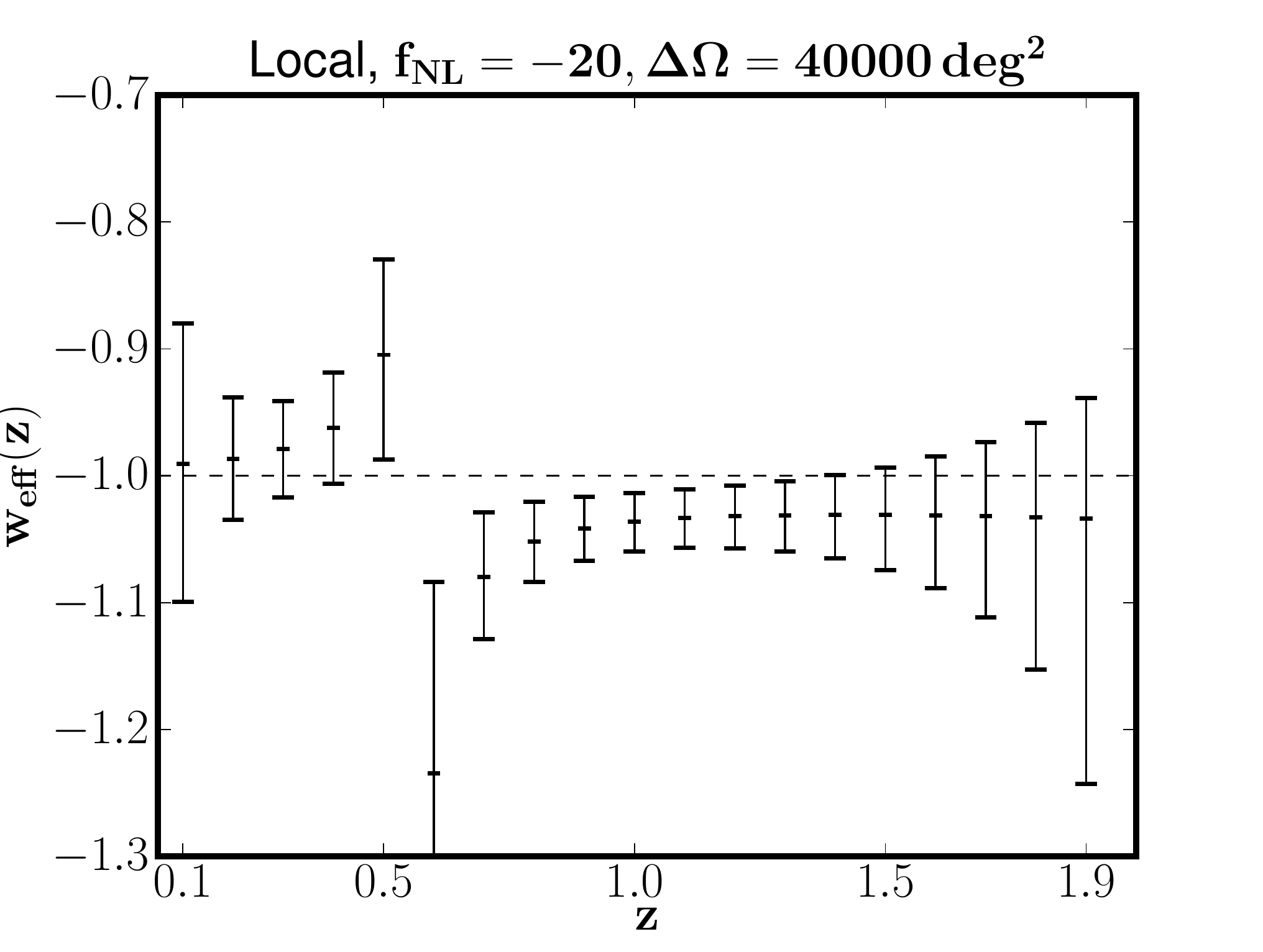}
\includegraphics[scale=0.42]{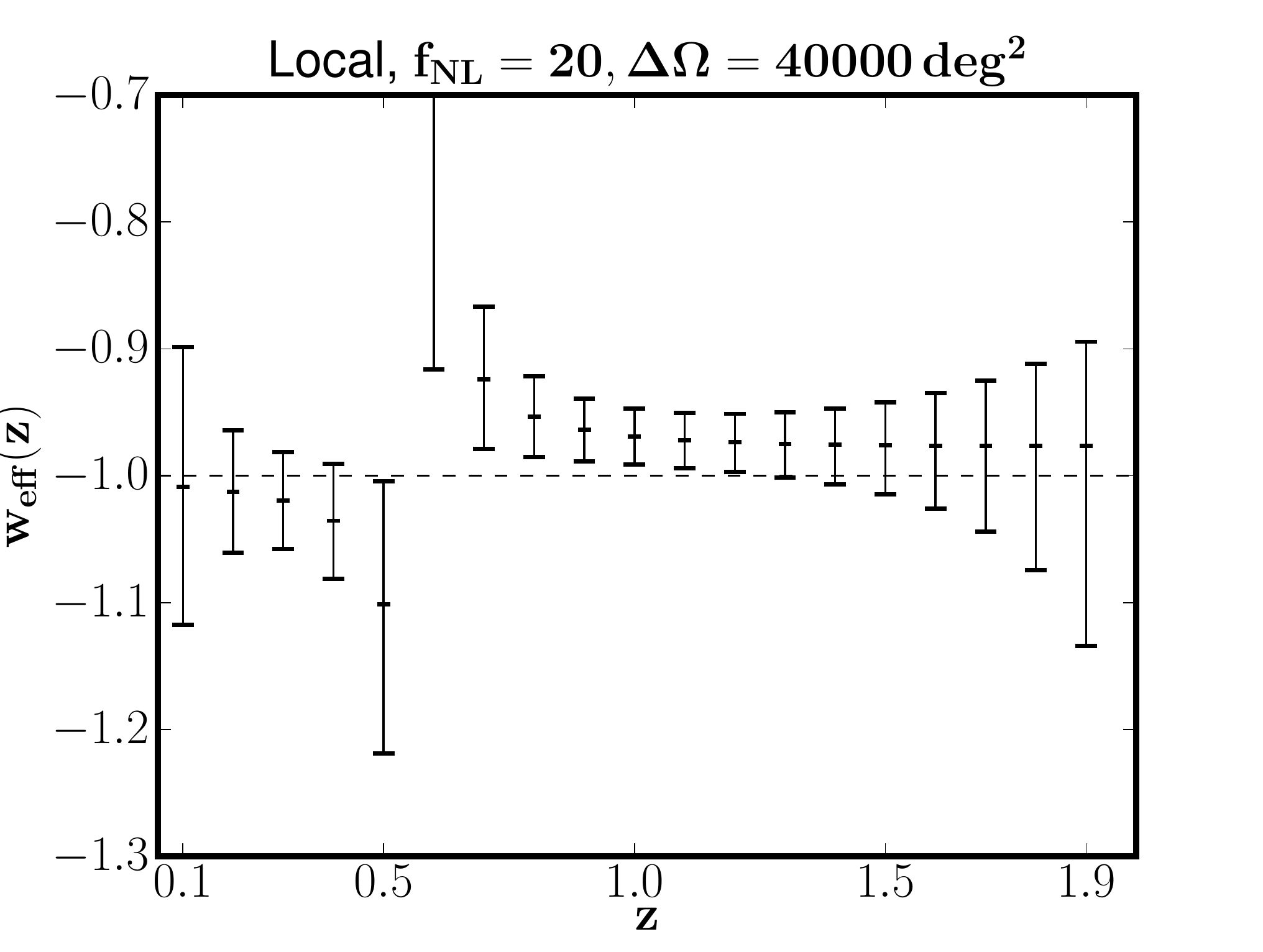}

\caption{The reconstructed effective dark energy equation of state for the Local parametrization with statistical observational uncertainties taken into account.}
\label{Figure4}
\end{figure*}

\begin{figure*}
\centering
\includegraphics[scale=0.42]{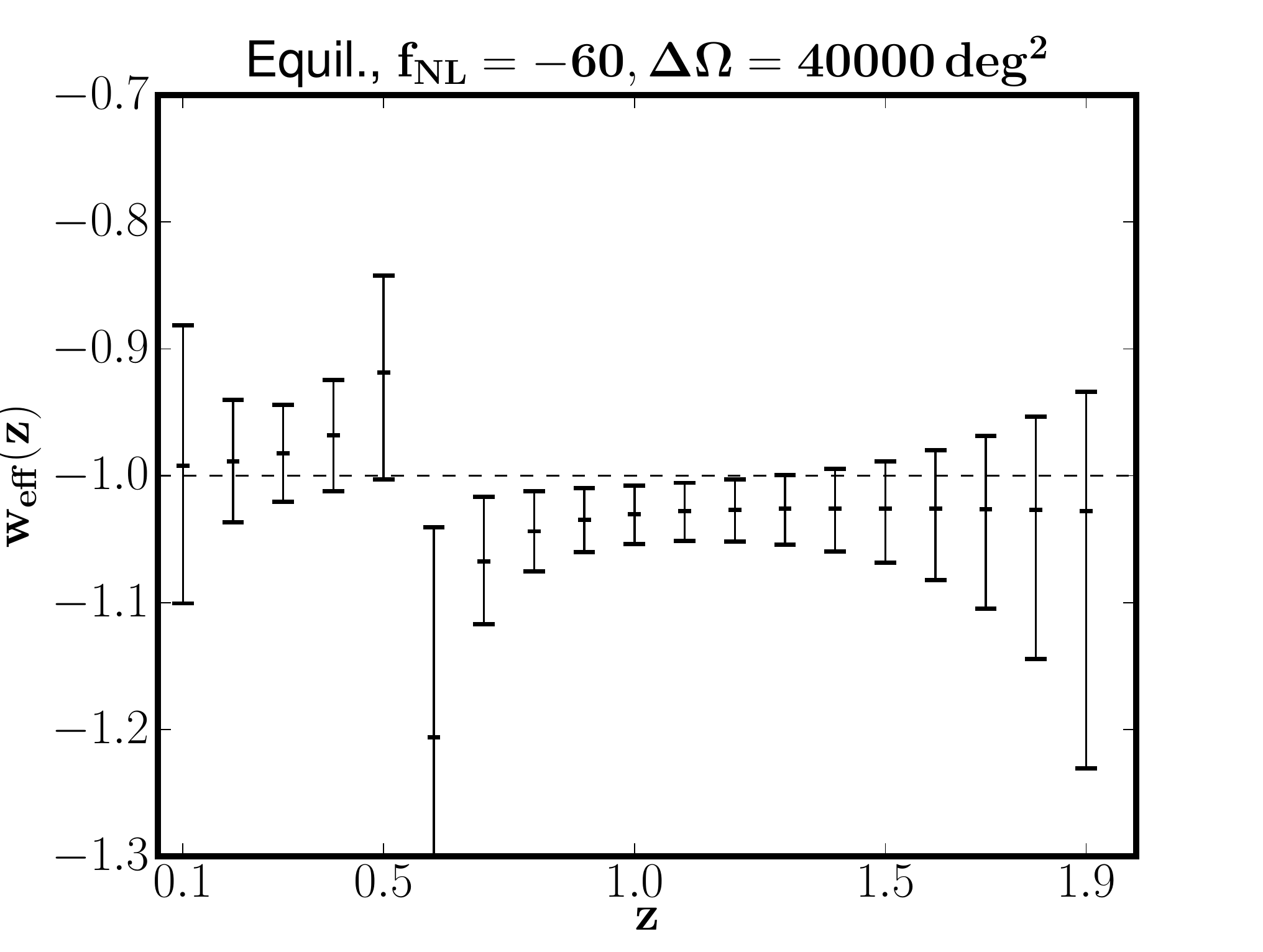}
\includegraphics[scale=0.42]{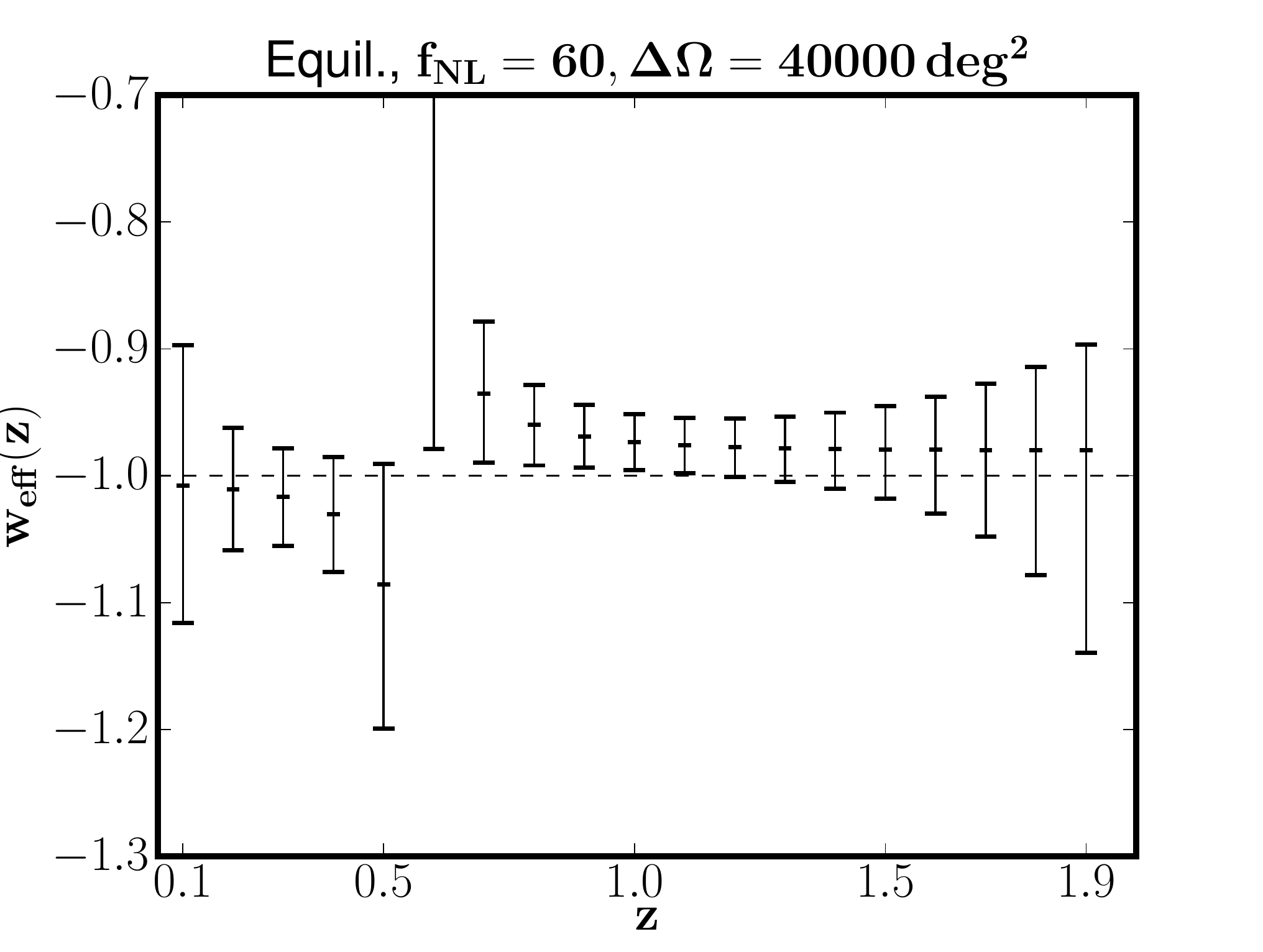}

\caption{The same as in Fig. \ref{Figure4}, but for the Equilateral parametrization.}
\label{Figure5}
\end{figure*}

Figures \ref{Figure4} and \ref{Figure5} show the $1\sigma$ statistical uncertainty in the reconstructed $w_{eff}$, as a function 
of redshift, when shot-noise is taken into account, and as before for a nearly full sky survey area of 40,000 square degrees. 
As can be seen, even when including the effect of such uncertainty, the apparent discontinuity on the evolution of $w_{eff}$ found 
in the previous subsection is still present (at least at the 2$\sigma$ confidence level) for values of $f_{NL}$ as small as $\pm20$ 
for the Local parametrization, and $\pm60$ for the Equilateral parametrization. Clearly, decreasing the survey area would 
mean that only values of $f_{NL}$ with larger modulus could be detected.

\vspace{-0.5cm}
\section{Conclusion}

In this work, we investigated whether the presence of primordial non-Gaussianities has an impact on the estimation 
of the effective dark energy equation of state, when one uses the abundance of galaxy clusters as a tool to probe
different cosmological scenarios. We computed the effective dark energy equation of state, $w_{eff}$ per redshift
bin, assuming Gaussian initial conditions, that is capable of reproducing the galaxy cluster counts expected in
several non-Gaussian models, thus constructing a correspondence $f_{NL}\,\mapsto\,w_{eff}$ for each redshift bin. 
The most important result of this work is the discovery of a redshift interval where no value for the effective dark
energy equation of state is capable of reproducing the non-Gaussian cluster abundance for models with $f_{NL}>0$, which is the result of there 
being a $w_{eff}$ that maximizes the cluster number density at each redshift, while for models with $f_{NL}<0$ a discontinuous
$w_{eff}$ is obtained. The appearance of such type of features may thus constitute a new diagnostic of the presence of primordial non-Gaussianities. 
Although, even under the ideal situation where only statistical uncertainties are present, their detection is only possible for values of $f_{NL}$ not too 
far away from those permitted by analysis of the WMAP 7-year data \citep{2011ApJS..192...18K}, it should be remembered that $f_{NL}$ may not be 
invariant with scale. In particular, there isn't any particular reason why $f_{NL}$ could not increase as the scale diminishes (see e.g. \citealt{2011PhRvD..83.41301R}).

\vspace{-0.5cm}
\section{Acknowledgements}

\small{
We thank Ant\'onio da Silva for useful discussions during the preparation of this paper and the anonymous referee for his comments. 
This work was partially funded by FCT (Portugal) through contract PTDC/CTE-AST/64711/2006 and FCOMP-01-0124-FEDER-015309 $\&$ PTDC/FIS/111725/2009. 
A. M. M. Trindade was supported by the FCT/IDPASC grant contract SFRH/BD/51647/2011.}

\vspace{-0.5cm}
{\small
\bibliographystyle{aa}

\begin{thebibliography}{}

\bibitem[{{Alishahiha} {et~al.}(2004){Alishahiha}, {Silverstein}, \&
  {Tong}}]{2004PhRvD..70l3505A}
{Alishahiha}, M., {Silverstein}, E., \& {Tong}, D. 2004, \prd, 70, 123505

\bibitem[{{Avelino} \& {Viana}(2000)}]{2000MNRAS.314..354A}
{Avelino}, P.~P. \& {Viana}, P.~T.~P. 2000, \mnras, 314, 354

\bibitem[{{Baldauf} {et~al.}(2011){Baldauf}, {Seljak}, {Senatore}, \&
  {Zaldarriaga}}]{2011arXiv1106.5507B}
{Baldauf}, T., {Seljak}, U., {Senatore}, L., \& {Zaldarriaga}, M. 2011, ArXiv
  e-prints

\bibitem[{{Bardeen} {et~al.}(1986){Bardeen}, {Bond}, {Kaiser}, \&
  {Szalay}}]{1986ApJ...304...15B}
{Bardeen}, J.~M., {Bond}, J.~R., {Kaiser}, N., \& {Szalay}, A.~S. 1986, \apj,
  304, 15
\bibitem[Wang et al.(2004)]{2004PhRvD..70l3008W} Wang, S., Khoury, J., 
Haiman, Z., \& May, M.\ 2004, \prd, 70, 123008 

\bibitem[{{Chiu} {et~al.}(1998){Chiu}, {Ostriker}, \&
  {Strauss}}]{1998ApJ...494..479C}
{Chiu}, W.~A., {Ostriker}, J.~P., \& {Strauss}, M.~A. 1998, \apj, 494, 479

\bibitem[{{Creminelli}(2003)}]{2003JCAP...10..003C}
{Creminelli}, P. 2003, \jcap, 10, 3

\bibitem[{{Fedeli} {et~al.}(2009){Fedeli}, {Moscardini}, \&
  {Matarrese}}]{2009MNRAS.397.1125F}
{Fedeli}, C., {Moscardini}, L., \& {Matarrese}, S. 2009, \mnras, 397, 1125

\bibitem[{{Fosalba} {et~al.}(2000){Fosalba}, {Gazta{\~n}aga}, \&
  {Elizalde}}]{2000ASPC..200..408F}
{Fosalba}, P., {Gazta{\~n}aga}, E., \& {Elizalde}, E. 2000, in Astronomical
  Society of the Pacific Conference Series, Vol. 200, Clustering at High
  Redshift, ed. {A.~Mazure, O.~Le F{\`e}vre, \& V.~Le Brun}, 408--+

\bibitem[{{Komatsu}(2010)}]{2010CQGra..27l4010K}
{Komatsu}, E. 2010, Classical and Quantum Gravity, 27, 124010

\bibitem[{{Komatsu} {et~al.}(2009){Komatsu}, {Dunkley}, {Nolta}, {Bennett},
  {Gold}, {Hinshaw}, {Jarosik}, {Larson}, {Limon}, {Page}, {Spergel},
  {Halpern}, {Hill}, {Kogut}, {Meyer}, {Tucker}, {Weiland}, {Wollack}, \&
  {Wright}}]{2009ApJS..180..330K}
{Komatsu}, E., {Dunkley}, J., {Nolta}, M.~R., {et~al.} 2009, \apjs, 180, 330

\bibitem[{{Komatsu} {et~al.}(2011){Komatsu}, {Smith}, {Dunkley}, {Bennett},
  {Gold}, {Hinshaw}, {Jarosik}, {Larson}, {Nolta}, {Page}, {Spergel},
  {Halpern}, {Hill}, {Kogut}, {Limon}, {Meyer}, {Odegard}, {Tucker}, {Weiland},
  {Wollack}, \& {Wright}}]{2011ApJS..192...18K}
{Komatsu}, E., {Smith}, K.~M., {Dunkley}, J., {et~al.} 2011, \apjs, 192, 18

\bibitem[{{Komatsu} \& {Spergel}(2001)}]{2001PhRvD..63f3002K}
{Komatsu}, E. \& {Spergel}, D.~N. 2001, \prd, 63, 063002

\bibitem[{{Lo Verde} {et~al.}(2008){Lo Verde}, {Miller}, {Shandera}, \&
  {Verde}}]{2008JCAP...04..014L}
{Lo Verde}, M., {Miller}, A., {Shandera}, S., \& {Verde}, L. 2008, \jcap, 4, 14

\bibitem[{{Lyth} \& {Rodr{\'i}guez}(2005)}]{2005PhRvL..95l1302L}
{Lyth}, D.~H. \& {Rodr{\'i}guez}, Y. 2005, Physical Review Letters, 95, 121302

\bibitem[{{Maldacena}(2003)}]{2003JHEP...05..013M}
{Maldacena}, J. 2003, Journal of High Energy Physics, 5, 13

\bibitem[{{Matarrese} \& {Verde}(2008)}]{2008ApJ...677L..77M}
{Matarrese}, S. \& {Verde}, L. 2008, \apjl, 677, L77

\bibitem[Matarrese et al.(2000)]{2000ApJ...541...10M} Matarrese, S., Verde, 
L., \& Jimenez, R.\ 2000, \apj, 541, 10 

\bibitem[{{Press} \& {Schechter}(1974)}]{1974ApJ...187..425P}
{Press}, W.~H. \& {Schechter}, P. 1974, \apj, 187, 425

\bibitem[{{Riotto} \& {Sloth}(2011)}]{2011PhRvD..83.41301R}
{Riotto}, A. \& {Sloth}, M.~S. 2011, \prd, 83, 041301(R)

\bibitem[{{Robinson} \& {Baker}(2000)}]{2000MNRAS.311..781R}
{Robinson}, J. \& {Baker}, J.~E. 2000, \mnras, 311, 781

\bibitem[{{Robinson} {et~al.}(2000){Robinson}, {Gawiser}, \&
  {Silk}}]{2000ApJ...532....1R}
{Robinson}, J., {Gawiser}, E., \& {Silk}, J. 2000, \apj, 532, 1

\bibitem[{{Salopek} \& {Bond}(1990)}]{1990PhRvD..42.3936S}
{Salopek}, D.~S. \& {Bond}, J.~R. 1990, \prd, 42, 3936

\bibitem[{{Seery} \& {Lidsey}(2005)}]{2005JCAP...06..003S}
{Seery}, D. \& {Lidsey}, J.~E. 2005, \jcap, 6, 3

\bibitem[{{Sefusatti} \& {Komatsu}(2007)}]{2007PhRvD..76h3004S}
{Sefusatti}, E. \& {Komatsu}, E. 2007, \prd, 76, 083004

\bibitem[{{Sheth} {et~al.}(2001){Sheth}, {Mo}, \&
  {Tormen}}]{2001MNRAS.323....1S}
{Sheth}, R.~K., {Mo}, H.~J., \& {Tormen}, G. 2001, \mnras, 323, 1

\bibitem[{{Slosar} {et~al.}(2008){Slosar}, {Hirata}, {Seljak}, {Ho}, \&
  {Padmanabhan}}]{2008JCAP...08..031S}
{Slosar}, A., {Hirata}, C., {Seljak}, U., {Ho}, S., \& {Padmanabhan}, N. 2008,
  \jcap, 8, 31

\bibitem[{{Sugiyama}(1995)}]{1995ApJS..100..281S}
{Sugiyama}, N. 1995, \apjs, 100, 281

\bibitem[{{Valageas} {et~al.}(2011){Valageas}, {Clerc}, {Pacaud}, \&
  {Pierre}}]{2011A&A...536..A95V}
{Valageas}, A., {Clerc}, C., {Pacaud}, U., \& {Pierre}, N. 2011,
  \aap, 536, A95
  
\bibitem[{{Wagner} {et~al.}(2010){Wagner}, {Verde}, \&
  {Boubekeur}}]{2010JCAP...10..022W}
{Wagner}, C., {Verde}, L., \& {Boubekeur}, L. 2010, \jcap, 10, 22

\end{thebibliography}

}

\end{document}